# Electrically Tunable Low Density Superconductivity in a Monolayer Topological Insulator

**Authors:** Valla Fatemi[1, #,], Sanfeng Wu[1, #, *], Yuan Cao[1], Landry Bretheau[2], Quinn D. Gibson[3], Kenji Watanabe[4], Takashi Taniguchi[4], Robert J. Cava[5], and Pablo Jarillo-Herrero[1, *]

**Affiliations:**
[1]Department of Physics, Massachusetts Institute of Technology, Cambridge, Massachusetts 02139, USA
[2]Laboratoire des Solides Irradiés, École Polytechnique, CNRS, CEA, 91128 Palaiseau Cedex, France
[3]Department of Chemistry, University of Liverpool, Liverpool L69 7ZX, United Kingdom
[4]Advanced Materials Laboratory, National Institute for Materials Science, 1-1 Namiki, Tsukuba 305-0044, Japan
[5]Department of Chemistry, Princeton University, Princeton, New Jersey 08544, USA
[#] These authors contributed equally to this work.
[*]Corresponding Email: swu02@mit.edu; pjarillo@mit.edu

**Abstract:** The capability to switch electrically between superconducting and insulating states of matter represents a novel paradigm in the state-of-the-art engineering of correlated electronic systems. An exciting possibility is to turn on superconductivity in a topologically non-trivial insulator, which provides a route to search for non-Abelian topological states. However, existing demonstrations of superconductor-insulator switches have involved only topologically trivial systems, and even those are rare due to the stringent requirement to tune the carrier density over a wide range. Here we report reversible, in-situ electrostatic on/off switching of superconductivity in a recently established quantum spin Hall insulator, namely monolayer tungsten ditelluride (WTe$_2$). Fabricated into a van der Waals field effect transistor, the monolayer's ground state can be continuously gate-tuned from the topological insulating to the superconducting state, with critical temperatures $T_c$ up to ~ 1 Kelvin. The critical density for the onset of superconductivity is estimated to be ~ $5\times10^{12}$ cm$^{-2}$, among the lowest for two-dimensional (2D) superconductors. Our results establish monolayer WTe$_2$ as a material platform for engineering novel superconducting nanodevices and topological phases of matter.

**One Sentence Summary:** Monolayer WTe$_2$, a quantum spin Hall insulator, exhibits superconductivity at very low electron densities.

A field effect switch between superconducting and insulating states allows, in-principle, tremendous possibilities for engineering novel superconducting devices. However, materials allowing for this switch reversibly and *in situ* are rare, and it remains difficult to implement such a device into existing technologies (*1*). An area where a suitable material may find unique application is at the intersection of superconductivity and topological insulators, which hosts a fertile landscape of interesting quantum phenomena, including non-Abelian topological excitations (*2–4*). Some topological insulators have been turned into superconductors via chemical doping (*5*, *6*), application of pressure (*7–9*), or via the proximity effect (*10*, *11*), which are either irreversible or *ex situ*. The last case and its analogs have recently attracted major efforts toward engineering Majorana physics, which has required fine-tuned interface engineering between distinct materials (*12–17*).

Here we report the observation of intrinsic superconductivity in a monolayer topological insulator, namely WTe$_2$, induced by the electric field effect. This monolayer transition metal dichalcogenide has recently been established as a quantum spin Hall insulator with robust edge transport up to 100 Kelvin (*18–23*). Our field effect device geometry (see Methods) is illustrated in Fig. 1A-B. The WTe$_2$ monolayer flake is van der Waals-encapsulated between two sheets of hexagonal boron nitride (BN) to protect it from chemical degradation and to improve transport characteristics. Top and bottom electrostatic gates, for which the same BN sheets serve as gate dielectrics, are fabricated to modulate the carrier density. An optical microscopy image of a typical device (device 1, BN thicknesses: 15 nm, top; 8 nm, bottom) is shown in Fig. 1B.

Figure 1C displays the temperature dependence of the four-probe resistance *R(T)* in device 1 when the monolayer is n-doped ($V_{bg}$ = 4V; $V_{tg}$ = 5V). *R* drops to zero at low temperatures from 1.2 kΩ in the normal state, and this zero-resistance state is present for a wide range of gate voltage parameters (Fig. 1C inset). The DC voltage-current (*V-I*) characteristics at various temperatures are shown in Fig. 1D, exhibiting the transition from Ohmic behavior (red curve) at ~ 1 K to highly nonlinear behavior (black curve) at low temperatures, including the characteristic zero voltage plateau for finite current, typical of superconductivity. This nonlinearity is further captured by the measurement shown in Fig. 1E, where the four-probe differential resistance, *dV/dI*, is plotted as a function of DC current bias $I_{DC}$ at base temperature. Under application of an out-of-plane magnetic field *B*, the nonlinear behavior is clearly suppressed (Fig. 1E), as expected for superconductivity. These observations confirm a superconducting state forming in the monolayer atomic sheet, in contrast to its 3D parent crystal, in which no superconductivity is found unless high pressure is applied (*24, 25*). We note that a small shoulder appears around 400 mK in the *R(T)* curve (Fig. 1C). While this may be an intrinsic feature of WTe$_2$, it could also be the result of having some inhomogeneous superconducting islands as the temperature is reduced. We then characterize the superconducting transition by quantifying several important temperatures. The onset of superconductivity, defined as the temperature at which *R* drops to 90% from its normal state, occurs at ~ 820 mK. Zero resistance is achieved at ~ 350 mK. Furthermore, we have also confirmed the 2D nature of the superconductivity by checking that the data fits well with the picture of a 2D Berezinskii-Kosterlitz-Thouless transition, with estimated critical temperature $T_{BKT}$ ~ 470 mK (Fig. S1). For simplicity in our discussion, we define the critical transition temperature $T_c$ as the temperature at which *R* equals 50% of the normal state value. For the curve in Fig. 1C, $T_c$ ~ 580 mK as indicated. Similar behavior is observed in device 2, in which $T_c$ is found to be as high as ~ 950 mK (Fig. S2, S3). Different from device 1, the *R(T)* characteristic of device 2 fits well with the standard Aslamazov-Larkin and Maki-Thompson fluctuation conductivity terms (*26, 27*) (Fig. S1).

The observed superconductivity, and hence the monolayer's ground state, is remarkably gate-tunable. We reveal this aspect by plotting $R(T)$ for different $V_{bg}$ (Fig. 2A), while $V_{tg}$ is kept fixed at 5 V. One can clearly see a critical gate voltage $V_c^{MIT} \sim -0.75$ V at which a metal to insulator transition occurs. At large gate voltage $V_{bg} > V_c^{MIT}$ (i.e. toward higher electron density), $R$ decreases with decreasing $T$, and a zero-resistance state is observed at low enough temperature, indicating a superconducting phase. At lower gate voltage, slightly above $V_c^{MIT}$, there may exist an intrinsic metallic (non-superconducting) phase at zero temperature, but a more detailed study is necessary to confirm this, as the observed behavior near the transition may also be consistent with the formation of both superconducting and non-superconducting regions as a result of density inhomogeneities. In contrast, when $V_{bg} < V_c^{MIT}$ (i.e. lower electron density), $R$ increases with decreasing $T$, pointing to an insulating ground state. This insulating behavior is consistent with the expectation for monolayer WTe$_2$ at low density: a topologically non-trivial insulating state exists at low temperatures (*20–22*), and edge state conduction is known to be marginal for the measurement configuration of device 1 (*20*). Figure S2 shows a similar transition behavior for device 2. Given that the field effect is applied through a moderate dielectric constant material ($\kappa_{BN} \sim 3.5$), these observations demonstrate that the ground state of the monolayer is unprecedentedly tunable between the two extremes of electron transport in materials, i.e. from the insulating to the superconducting state. For comparison, existing superconducting systems showing similar field effect behavior are achieved only by extreme charge density doping using ionic liquids (*28*), ultrahigh-$\kappa$ dielectric materials such as SrTiO$_3$ (*29, 30*), or ferroelectric polarization (*31*).

The gate tunability of the superconducting state can be further characterized by the extraction of the critical temperature and measurements of the critical current. In the upper panel of Fig. 2B, we plot the extracted $T_c(V_{bg})$, where one finds increasing $T_c$ with increasing $V_{bg}$. Extrapolating to $T_c = 0$ K (red curve, see also Fig. S4), we find a critical gate voltage $V_{c1} \sim -0.65$ V. The maximum observed $T_c \sim 0.61$ K appears at the highest applicable gate voltages for this device. Fig. 2C shows $dV/dI$ vs $I_{DC}$ at $V_{bg} = V_{tg} = 5$ V, where the sharp peak at the critical current $|I_{DC}| = I_c$ and the zero-resistance plateau for low currents demonstrate the expected behavior for superconductivity. The additional peak at $|I_{DC}| = I_c'$ in this data set is consistent with the previously mentioned small shoulder in the $R(T)$ characteristic (Fig. 1C). The differential resistance is strongly modulated by gate voltage, as shown in Fig. 2D. The peaks at $\pm I_c$ monotonically shift towards zero bias with decreasing $V_{bg}$ and eventually merge into a single peak at zero bias in the insulating region. This modulation of $I_c$ is summarized in the lower panel of Fig. 2B, where we find that the critical gate voltage $V_{c2}$, at which $I_c$ vanishes, falls in the range between -0.6 V and -0.8 V.

Therefore, one finds that $V_{c1}$, $V_{c2}$, and $V_c^{MIT}$ are very close to each other, which identifies the critical gate voltage as $V_{bg} = V_c \sim -0.7$ V (when $V_{tg} = 5$ V), above which superconductivity is exhibited. Using the capacitance model, we estimate the corresponding critical doping density as $n_c \sim 5 \times 10^{12}$ cm$^{-2}$. Hall effect measurements confirm a low absolute carrier density in the superconducting region which is consistent with the density estimates from the electrostatic capacitance model (Fig. 2E-F and details in Fig. S5). The visible deviation appearing at lower doping in the plot may be related to inhomogeneous transport in the monolayer (e.g. the presence of conducting edge channels), which can introduce a minor factor to the measured Hall density. A similar critical density (~ $3\times10^{12}$ cm$^{-2}$) is found in device 2. These are among the lowest critical density values reported for 2D superconductors (*32*). Such low-density superconductivity in monolayer WTe$_2$ is the key to its extreme gate tunability. We also note that the maximum $T_c$ obtained here (~0.6 K for device 1 and ~1 K for device 2, at carrier density $n_{2D}$ ~ $1.8\times10^{13}$ cm$^{-2}$) is relatively high, compared to other 2D superconductors with similar density (*30*, *32*).

The preliminary low temperature electronic phase diagram of monolayer WTe$_2$ can now be summarized, as shown in Fig. 3A. Near zero charge density resides the quantum spin Hall insulator (QSHI) phase, robust up to 100 K (*23*). With n-type doping above the critical density, $n_c$, the superconducting ground state develops at temperatures below $T_c$ (superconductivity was not observed with p-type doping in the gate-accessible region). Above $T_c$, the system is metallic. The coexistence of QSHI and superconducting states in the same phase diagram establishes monolayer WTe$_2$ as a unique material for observing physics at the intersection of topological insulating states and superconductivity. In Fig. 3B, we show data from device 2, in which superconductivity and the quantized QSH transport are observed in the same monolayer device. The helical edge transport in the QSHI phase in this device was characterized in detail in our previous report (*23*). Its superconducting behavior is characterized in Fig. S3. These results therefore demonstrate the electrostatic field effect switching of superconductivity on and off in a QSHI system. Moreover, in the same device we also observed preliminary evidence that superconductivity can be introduced into the helical edge states by proximity effect via adjacent gate-induced superconducting regions (see details in Fig. S6). In principle, such proximity-induced superconducting helical edge states can be used to construct devices hosting Majorana zero modes to study non-Abelian physics.

We further characterize the gate-tunable superconducting properties by examining magnetic field dependence. As shown Fig. 4A, one finds an out-of-plane critical field at base temperature $B_{c2,\perp}$ ~ 30 mT (for $V_{tg} = V_{bg} = 5$ V), at which half of the normal state resistance is recovered. The temperature and gate dependence of this critical field is summarized in Fig. S7, which indicates that the Ginzburg-Landau coherence length is close to 100 nm. This length is about an order of magnitude larger than the estimated transport mean free path (Fig. S4), suggesting that the superconductivity is in the dirty limit. In contrast to the out-of-plane critical field, the in-plane critical field $B_{c2,//}$ is significantly larger: about 4.3 T is required to recover half the normal state resistance, as shown in Fig. 4B. This field is about 4 times the conventional Pauli paramagnetic limit, which is given by $1.84T_c$ ~ 1.1 T. We summarize the $B_{c2}$ - $T_c$ phase diagram for both the in-plane and out-of-plane cases in Fig. 4C. We note that the crystal symmetry and band structure of monolayer WTe$_2$ are very different from the 2H-type transition metal dichalcogenides, in which Ising-type superconductivity is responsible for exceeding the Pauli limit (*32*). Other possible mechanisms for exceeding the Pauli limit include reduced electron g-factor, spin-triplet pairing, or strong spin-orbit scattering (*33*, *34*). The superconductivity in the dirty limit is consistent with the spin-orbit scattering scenario (*33*, *34*), in which we extract a spin-orbit scattering time ~ 244 fs

(see fit in Fig. 4 and details in the supplementary materials). However, we stress that our observations merit further study to understand the nature of the superconductivity and the exact mechanism for the large in-plane critical field in the WTe$_2$ monolayer.

We comment now on several interesting directions to explore. One is to look for an optimal doping, e.g. to determine whether a superconducting dome exists in the phase diagram (*24*, *30*, *35*). Another is to investigate the effects of the material's strong anisotropy (*36*) and tunable non-centrosymmetry (*18*, *37*) on the superconductivity. A particularly interesting question is whether the observed superconductivity is topologically non-trivial. Although we currently do not have an answer, our results point to an exciting possibility of creating a 2D-crystal based topological superconductor using the proximity effect. Indeed, using local gates to achieve lateral modulation of superconducting and QSHI regions would enable the fabrication of tunable superconductor-QSHI-superconductor topological Josephson junctions (Fig. S6) and the investigation of Majorana modes in a single material. In addition, van der Waals heterostructures which interface monolayer WTe$_2$ with materials such as the recently discovered 2D layered ferromagnets (*38*, *39*) can be developed for studying the interplay between superconductivity, magnetism, and topology.

**Acknowledgments:**

We thank Liang Fu for helpful discussions and Joel I-Jan Wang for assistance in operating the dilution refrigerator. This work was partly supported through AFOSR Grant No. FA9550-16-1-0382 as well as the Gordon and Betty Moore Foundation's EPiQS Initiative through Grant No. GBMF4541 to P. J-H. Device nanofabrication was partly supported by the Center for Excitonics, an Energy Frontier Research Center funded by the DOE, Basic Energy Sciences Office, under Award No. DE-SC0001088. This work made use of the Materials Research Science and Engineering Center's Shared Experimental Facilities supported by NSF under Award No. DMR-0819762. Sample fabrication was performed partly at the Harvard Center for Nanoscale Science supported by the NSF under Grant No. ECS-0335765. S.W. acknowledges the support of the MIT Pappalardo Fellowship in Physics. The WTe$_2$ crystal growth performed at Princeton University was supported by an NSF MRSEC grant, DMR-1420541. Growth of BN crystals was supported by the Elemental Strategy Initiative conducted by the MEXT, Japan and JSPS KAKENHI Grant Numbers JP15K21722 and JP25106006.

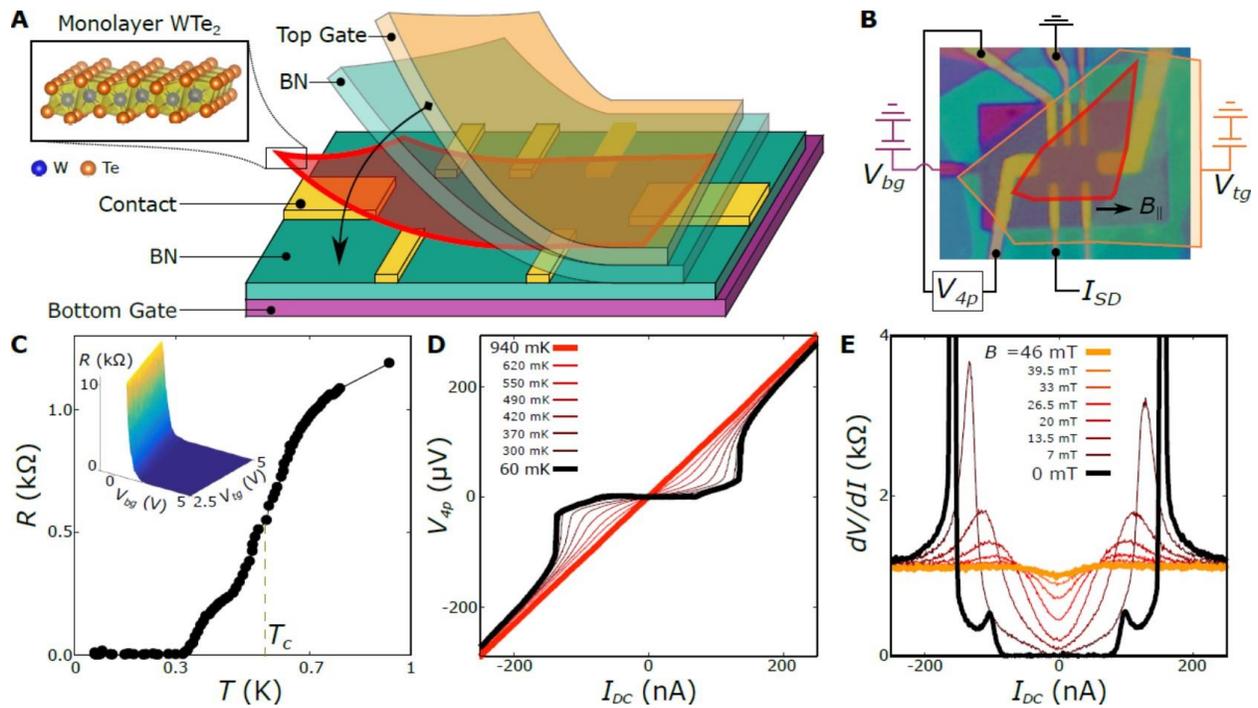

**Figure 1 | Device schematic and superconductivity characteristics.** (**A**) Cartoon illustration of the device structure and the crystal structure of monolayer $WTe_2$. (**B**) Optical microscopy image of device 1, with the monolayer $WTe_2$ (red) and graphite top gate (orange) highlighted. Circuit elements show the measurement configuration. (**C**) Temperature dependence of the resistance for $V_{bg}$ = 4 V and $V_{tg}$ = 5 V. Inset: the resistance as a function of both gate voltages, at base temperature. (**D**) *V-I* characteristics from base temperature (black) up to 940 mK (red). (**E**) Nonlinear *V-I* behavior, captured by differential resistance curves, for different perpendicular magnetic fields at base temperature.

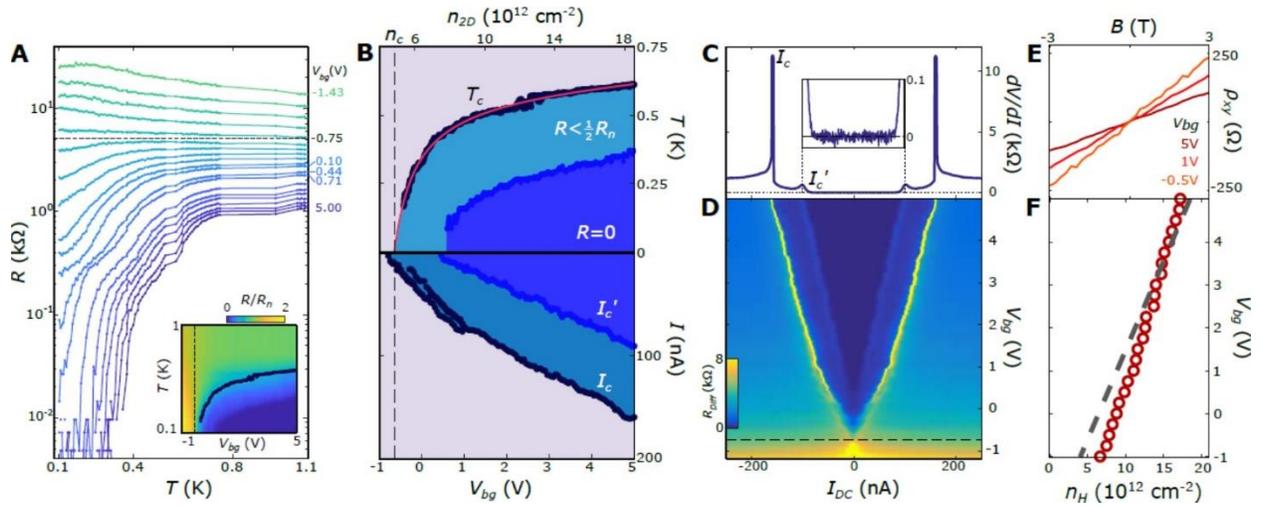

**Figure 2 | Switching on/off superconductivity with an electrostatic gate.** (**A**) $R(T)$ characteristic for different gate voltages, showing the transition from a superconducting state to an insulating state. Dashed line is a guide to the gate voltage, $V_c^{MIT}$, that separates the two regimes. Inset: color plot of the same data, normalized to the normal state resistance $R_n$, with $T_c$ marked in black. (**B**) Upper panel: gate dependent critical temperature $T_c$, summarized from **A**. The zero-resistance region is shaded dark blue. Lower panel: gate dependent critical current, $I_c$ and $I_c'$, summarized from **D**. For $V_{bg} < 1V$, $I_c$ is found to bifurcate into two peaks, the values of which are determined by extrema of the second derivative of $dV/dI$ ($I_{DC}$) data. The corresponding electron density ($n_{2D}$), estimated from the capacitance model, is shown on the top axis. (**C**) Differential resistance $dV/dI$ v.s. current bias $I_{DC}$ for $V_{tg} = V_{bg} = 5V$. Inset: zoom-in to the zero-resistance region. (**D**) Differential resistance $dV/dI$ vs. current bias $I_{DC}$ and gate voltage $V_{bg}$. Close to the $V_{bg} \sim 1$, the observed $I_c$ trace bifurcates, as more clearly indicated in **B**. Dashed line indicates the gate voltage at which the peaks in differential resistance merge at zero bias, indicating destruction of superconductivity. (**E**) Hall resistivity $\rho_{xy}(B)$ for three selected bottom gate voltages taken at 2 K ($V_{tg} = 5V$). See supplementary materials for details. (**F**) Extracted Hall density as a function of the gate voltage (red circles) and the estimated density from the capacitance model (grey dashed line).

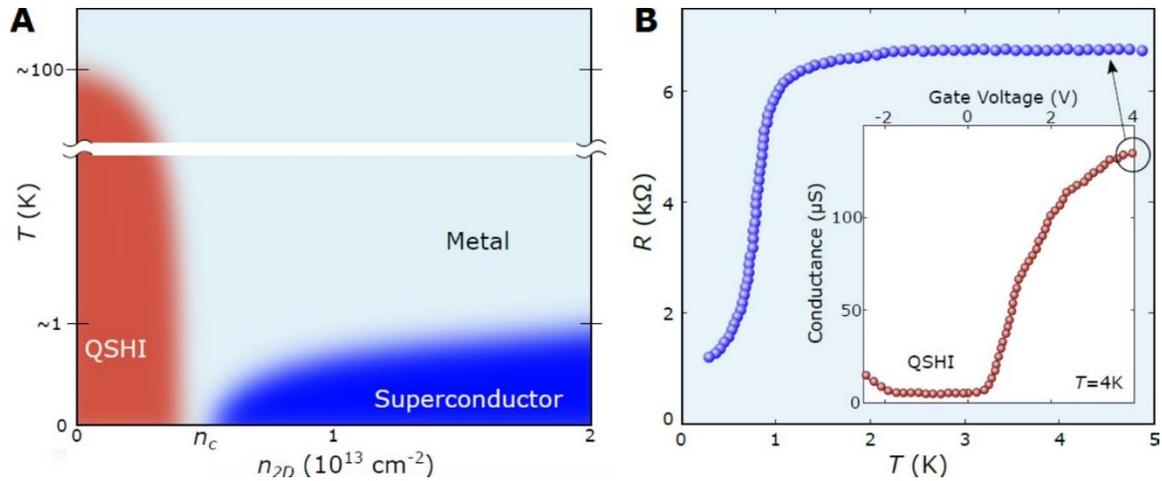

**Figure 3 | Electronic phase diagram of monolayer WTe$_2$.** (**A**) Schematic temperature-density phase diagram for the electronic ground state of monolayer WTe$_2$, reminiscent of the inset of Fig. 2A. (**B**) Resistance *vs.* temperature for device 2, which again shows a clear superconducting transition. The residual resistance at low temperature is due to a known imperfection in this device, as described in the supplementary materials. *Inset*: gate-dependent conductance of device 2, where the observed plateau corresponds to the QSHI phase.

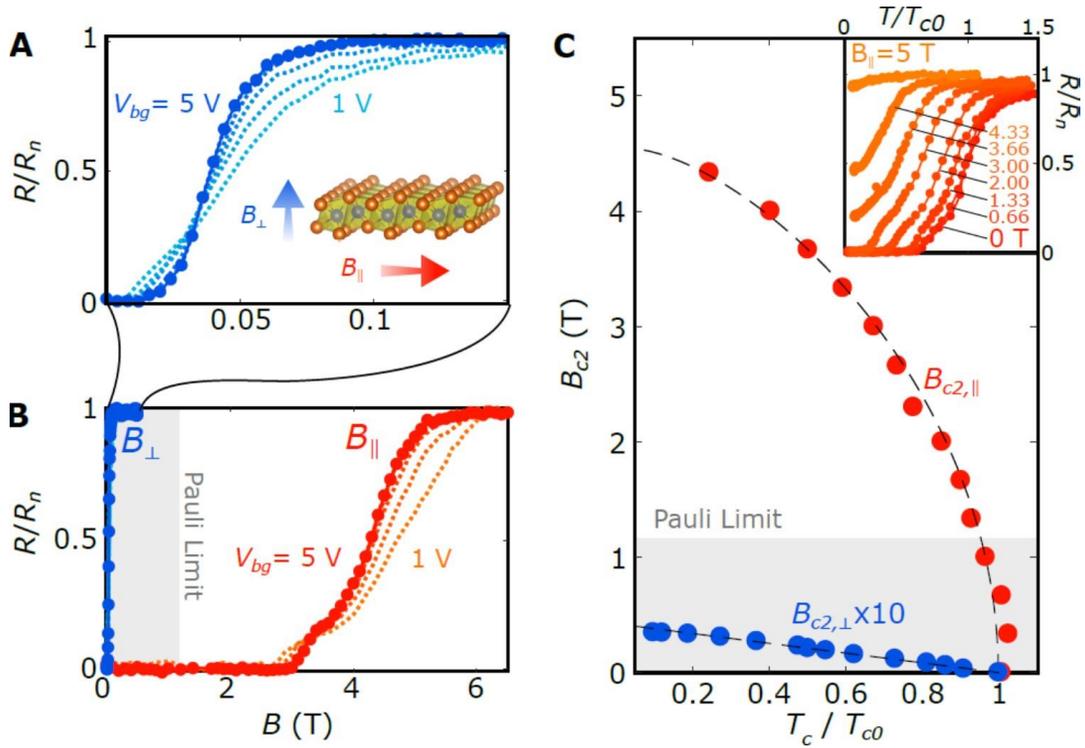

**Figure 4 | Magnetic field effect on the monolayer superconductivity.** (**A**) Perpendicular magnetic field dependence of the resistance, $R(B_\perp)$, normalized to the normal state resistance $R_n$, for different gate voltages. Inset: schematic of the perpendicular and parallel field orientations. See Fig. 1B for orientation of parallel magnetic field in the plane of the device. (**B**) Orange: parallel magnetic field dependence of resistance, $R(B_\parallel)$, normalized to the normal state resistance. Blue: perpendicular magnetic field dependence plotted on the same scale for comparison. The data is from Device 1 with $V_{tg} = 5$V at base temperature. (**C**) $B_{c2}$ - $T_c$ phase diagram for both the parallel (orange) and perpendicular (blue) orientations. $V_{tg} = V_{bg} = 5$V. Black dashed lines are fits to theoretical models (see supplementary materials). Inset: the temperature dependence of resistance under different parallel magnetic fields. Data lines in A, B, and inset of C are regularly spaced between the labeled extremal values.

## Supplementary Materials for

**Electrically Tunable Low Density Superconductivity in a Monolayer Topological Insulator**

**Authors:** Valla Fatemi[1, #], Sanfeng Wu[1, #, *], Yuan Cao[1], Landry Bretheau[2], Quinn D. Gibson[3], Kenji Watanabe[4], Takashi Taniguchi[4], Robert J. Cava[5], and Pablo Jarillo-Herrero[1, *]

[#]These authors contributed equally to this work.

[*]Correspondence to: swu02@mit.edu; pjarillo@mit.edu

**This PDF file includes:**

Materials and Methods

Supplementary Text

Figs. S1 to S7

**Materials and Methods**

Device Fabrication

The WTe$_2$ bulk crystals were grown as described in a previous work (*1*). The electrostatic gates and contacts are fabricated prior to the exfoliation of WTe$_2$ utilizing standard electron beam lithography and metal vapor deposition techniques. The exfoliation and encapsulation of WTe$_2$ monolayers is accomplished in an argon atmosphere glove box to avoid degradation. Details of such a pre-contact fabrication procedure were described in Ref (*2*).

Electrical Measurements

Device 1 was cooled in a dilution fridge equipped with electronic filters, described in a previous work (*3*), to obtain an electron temperature of ~ 100 mK. Device 2 was measured in a $^3$He fridge with base temperature ~ 300 mK. Transport measurements are conducted with a lock-in technique with low-frequency (between 5 and 17 Hz) AC current excitation (~ 1 nA).

**Supplementary Text**

Temperature Dependence Analysis

In Fig. S1 we inspect temperature dependent resistance *R(T)* for device 1 and 2 with a representative gate voltage configuration (the same data in main text; Fig. 1C for device 1 and Fig. 3B for device 2). In Fig. S1A-B we fit the *R(T)* characteristic with the Aslamazov-Larkin and Maki-Thompson fluctuation conductivity terms (*4–7*), which is often used to describe the superconducting transition above $T_c$ in the dirty limit. The analysis of mean free path and the coherence length can be found below (Figs. S4 and S7). While this model indeed fits to our data taken from device 2 at the high temperature side (with extracted mean-field transition temperature ~ 0.79 K), it fails to capture the data taken from device 1. We point out that existing models also fail to capture the observed superconductivity in other monolayer systems, such as NbSe$_2$ and TaS$_2$ (*8*, *9*). We hope that these observations can stimulate research to better characterize the superconducting transition of crystalline monolayer superconductors.

In Fig. S1C-D we perform Berezinskii-Kosterlitz-Thouless (BKT) analysis for device 1. The nonlinear *V-I* curves (inset in Fig. S1C, the same data as Fig. 1D) can be characterized by a power-law dependence $V_{4p} \propto I_{DC}^\alpha$ near the critical current $I_c$, where $\alpha$ is the exponent. $\alpha$ increases rapidly from unity at high temperature (near 1 K) as temperature is lowered. The BKT transition temperature, at which $\alpha = 3$, is $T_{BKT} = 470$ mK. According to BKT theory, below this critical temperature, the resistance should evolve as: $\ln(R'(T)/R_0) = - C (T/T_{BKT} - 1)^{-1/2}$, where *C* and $R_0$ are constants. Fig. S1D plots the measured resistance from Fig. S1B (with an offset to compensate for the small shoulder, $R' = R - 250$ Ω; $R_0 = R_n$, the normal state resistance) in log scale as a function of the reduced temperature $(T/T_{BKT} - 1)^{-1/2}$, which shows the expected exponential behavior. Therefore, this data is consistent with BKT transition for 2D superconductors.

Gate dependence data in device 2

The gate dependent behavior of the observed superconductivity in device 2 is shown in Fig. S2. Like the situation in device 1 (Fig. 2 in the main text), here we observe a continuous, reversible tuning from superconducting state all the way to the (topological) insulating state. The transition temperature in this device approaches 1 K, which is higher than in device 1. This same device was previously investigated in the QSHI regime and exhibited helical edge transport (*2*), which confirms that superconductivity and QSHI can be observed in the same device. Moreover, different regions can be locally tuned into the QSHI and superconducting regimes to form a lateral junction for engineering proximity-induced superconducting helical edge states (see below and Fig. S6).

Details of superconductivity in device 2

In Fig. S3 we describe the observed superconductivity in device 2. Fig. S3A illustrates the device design, the details of which are reported in Ref. (*2*). The device was measured in a $^3$He fridge with base temperature of 0.3 K. At base temperature and finite electron doping, superconductivity is observed in the region where the large portion of the monolayer WTe$_2$ is fully encapsulated by BN, i.e., from the left-side contacts across the device to the right-side contacts. In contrast, no signature of superconductivity is found in the left sides or right sides separately, where the bottom face of WTe$_2$ is not protected by BN due to suspension by contacts (half encapsulation). The situation is summarized in Fig. S3 B, C, and D (at $V_{tg}$ = 4 V). We note that for the data shown in Fig. S3C, there is also a portion of WTe$_2$ between the contacts that is half encapsulated, due to suspension by the inner-most contacts. As a result, we do not observe zero resistance in this device since the half-encapsulated regions always contribute a finite resistance. In Fig. S3 E and F, we characterize the superconductivity by measuring the differential resistance as a function of applied DC current and perpendicular magnetic field at base temperature, which shows behavior similar to Device 1.

Relation between $T_c$ and $R_n$ and the extrapolation of critical gate voltage

Figure S4 shows the relation between $T_c$ and $R_n$ for device 1, which is much less non-linear than the relationship between $T_c$ and $V_{bg}$. Also shown are fits to a phenomenological cubic relationship between $T_c$ and $R_n$ (red) and a linear relationship for low $T_c$ (blue). Because we can measure $R_n$ for a range of gate voltages extending beyond the range at which $T_c$ is measurable, we can extrapolate $T_c$ based on these fits, as shown. While the extended cubic fit is used in Fig. 2B of the main text, we note that the linear (blue) and cubic (red) fittings yield similar critical gate voltages, as indicated in the figure.

Estimation of mean free path

The mean free path can be estimated based on knowledge of the carrier density $n_{2D}$, total band degeneracy $g$, and conductivity: $\sigma = k_F l_{mfp} \frac{g}{2} \frac{e^2}{h}$ (an isotropic system is assumed here for simplicity), where $g$ is assumed to be 4 counting for both spin and valley. We use finite element analysis to estimate the geometry factor of 0.47 to convert the measured conductance (see next section) to the conductivity (this is equivalent to a length-to-width ratio for a Hall bar geometry). The Fermi wavevector is given by $k_F = \sqrt{4\pi n_{2D}/g}$. As a result, we find $l_{mfp}$ is several nanometers for all gate voltages, as shown in the inset to Fig. S4.

Estimation of electron doping density

We first estimate the carrier density in the monolayer based on an electrostatic capacitance model. The density is $n = \frac{\kappa \varepsilon_0}{e}\left(\frac{V_{tg}}{d_t} + \frac{V_{bg}}{d_b}\right)$, where $\varepsilon_0$ is the vacuum permittivity, $\kappa \sim 3.5$ is the relative dielectric constant of thin BN, $e$ is the electron charge, and $d_t$ ($d_b$) is the thickness of the top (bottom) BN. For device 1 the BN thicknesses are $d_t = 15$ nm and $d_b = 8$ nm, and for device 2 they are $d_t = 9$ nm and $d_b = 13$ nm. The critical gate voltage for superconductivity in device 1 is $V_{bg} \sim -0.7$ V at $V_{tg} = 5$ V, which results in $n_c \sim 4.7\times10^{12}$ cm$^{-2}$. The critical density is $\sim 3\times10^{12}$ cm$^{-2}$ for device 2.

We also performed Hall measurements on Device 1 (Fig. S5A) for the measurement configuration shown in Fig. S5B. Due to the irregular shape of the monolayer and the closely spaced contacts, a geometric factor needs to be considered for extracting the true Hall density. We performed numerical simulations with COMSOL Multiphysics (see Fig. S5B for the simulated device configuration) and obtain the geometry factor $\sim 2$, which is accounted for in the antisymmetrized data in Fig. S5C as our extraction of the Hall resistivity $\rho_{xy}$. The accuracy of the simulation is independently checked by looking at the mixing of the $V_{xy}$ and $V_{xx}$ signals at zero magnetic field: the experimentally measured ratio $V_{xy}/V_{xx}$ ranges between -0.8 and -1.0 for nearly the entire range of accessible gate voltages, which agrees well with the simulated value of -0.9 (Fig. S5D). The Hall density, after considering the geometry factor, as a function of gate voltage is presented in Fig. S5E (same as in Fig. 2F), which, in the highly doped regime, exhibits good agreement with the capacitance model discussed above. The deviation appearing in the lower doping region may be related to inhomogeneous transport in the monolayer (e.g. the presence of conducting edge channels), which can cause the measured Hall effect density to be less accurate.

We also note that the device exhibits an ambipolar field effect characteristic centered near zero gate voltage (Fig. S5F). This suggests that monolayer WTe$_2$ is undoped at zero gate voltage and that the gate introduces electron-type or hole-type doping dependent on its sign. The situation is very similar to graphene (a 2D material with semi-metallic behavior and ambipolar gate dependence), in which systematic investigations confirm that the capacitance model agrees with the true density.

Proximity induced superconducting helical edge state in device 2

In Fig. S6, we describe preliminary experimental evidence for a *proximity effect in the helical edge mode*, realized in one of our reported devices (Device 2). Such proximity-induced superconducting helical edge modes, if confirmed, would constitute the long-sought 1D time-reversal invariant topological superconductor and naturally host Majorana modes. Our Device 2 can already realize this interesting situation by combining the top and bottom gates (Fig. S3A). We first set the top gate voltage at a high value (3.5V) and then use one of the local bottom gates to deplete the carriers in the middle of the flake with a length $\sim$ 100nm. Total differential resistance of the monolayer flake as a function of source to drain bias and the selected local bottom gate is shown in Fig. S6A. The monolayer is highly doped so that it will become superconducting at low temperature, except in the locally depleted region, where the QSH insulating state exists. The two situations (entirely doped flake *vs* doped-QSH insulator-doped junctions in monolayer) are illustrated by the cartoon pictures. The data was taken at 300 mK. The line cuts indicated by the arrows in Fig. S6A are shown as the inset of Fig. S6B, where the bottom gate is set to 2V (yellow), -5.8V (red) and -6.8V (blue). The red and blue curves have higher resistance due to the presence of the QSH insulating region and their differences with the yellow curve measure the resistance of the QSH edge channel (see details in Ref. (*2*)), which is plotted as the main panel in Fig. S6B. At high DC bias, the superconducting state is lost, and the measured edge state resistance is a bit higher than the ideal value of $h/2e^2$ for two parallel edges, which is normal for a QSH state (*2*, *10*). However, at zero bias, where the superconductivity in the doped region is formed, the edge channel yields clearly a resistance that is *smaller* than the expected quantized value of $h/e^2$ per edge, which is never observed when the contact to the QSH edge is a normal-state metal (*2*, *10*). This is clear evidence that the superconducting proximity effect to the helical edge mode of the QSH state can be accomplished by a lateral junction within the monolayer. Efforts can now be directed to optimize the device structure and the measurement scheme to realize proximity induced topological superconductivity in the helical edge states and explore the physics of Majorana modes.

Perpendicular magnetic field effect and coherence length

The superconducting transition under perpendicular magnetic field has strong gate dependence, as shown in Fig. S7A, where the transition becomes sharper and slightly shifted towards lower field at higher doping. The contour corresponding to $B_{c2,\perp}$ (defined here by $R/R_N = 0.5$) is marked in the $R(B, V_{bg})$ color map (Fig. S7B). Our data suggests that the critical field and critical temperature may have opposite gate dependence. The temperature dependence of $B_{c2,\perp}$ is summarized in Fig. S7C. At fixed $V_{bg}$, the data is well captured by the 2D Ginzburg-Landau theory which predicts $B_{c2,\perp}(T) = \Phi_0/[2\pi\xi_{GL}(0)^2] [1-T/T_c(0)]$, where $\Phi_0$ is the magnetic flux quantum, $T_c(0)$ is the transition temperature at zero field, and $\xi_{GL}(0)$ is the coherence length at zero temperature assuming in-plane isotropy. Figure S7D plots the extracted $\xi_{GL}(0)$, which suggests that its value may slightly increase with increasing $V_{bg}$, reaching ~ 90 nm at 5 V. That the coherence length increases despite increasing $T_c$ is indicative of higher Fermi velocity at higher density.

Fitting procedure for parallel critical field vs temperature

In the main text, we have reported that the parallel critical magnetic field $B_{c2,//}$ exceeds the Pauli limit significantly. While the exact mechanism for this to happen is yet determined, here we fit the experimental data with an existing model considering spin-orbit scattering from impurities. The relationship between critical temperature and magnetic pair breaking is given by

$$\ln\frac{T_c}{T_{c0}} = \psi\left(\frac{1}{2}\right) - \psi\left(\frac{1}{2} + \frac{\alpha}{2\pi k_B T_c}\right) \tag{1}$$

where $\alpha$ is the pair breaking parameter that depends on the applied magnetic field (*11*, *12*). The enhancement of the parallel critical field compared to the Pauli limit may come from spin-orbit scattering, in which case we have

$$\alpha = \frac{g^2 \mu_B^2 B_{c2,\parallel}^2}{2\hbar \tau_{so}^{-1}} \qquad (2)$$

where $\tau_{so}$ is a spin-orbit scattering time, $g$ here is the electronic g-factor, and $\mu_B$ is the Bohr magneton (11, 12). This is valid when $\hbar \tau_{so}^{-1} \gg g\mu_B B_{c2,//}$. By fitting the extracted $T_c$ and $B_{c2,//}$ to the above relationships (data and fit are shown in Fig. 4C of the main text), we obtain $\tau_{so} = 244 fs$ for $V_{bg} = V_{tg} = 5$V. This gives $\hbar \tau_{so}^{-1} \approx 2.7 meV \approx 5 g \mu_B B_{c2,\parallel}$ (assuming $g = 2$), which is consistent with the use of this expression.

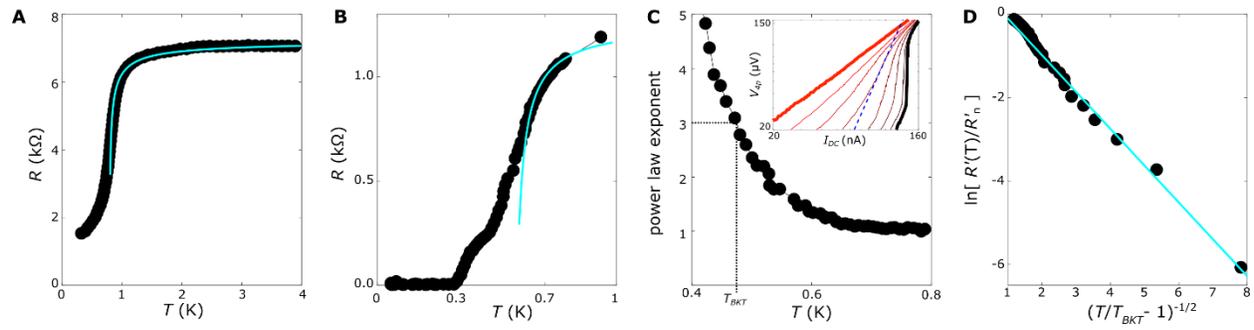

**Fig. S1 Transition Temperature Analysis.** **(A-B)** Fluctuation conductivity fit (utilizing the Aslamazov-Larkin and Maki-Thompson terms) for the high-temperature part of the superconducting transition in Device 2 and 1, respectively. Mean field temperatures of $T_c = 0.59$ K and 0.77 K are extracted. **(C)** Power law exponent of the V-I curve as a function of temperature. Inset: log-log plot of sample V-I traces for different temperatures (solid lines) with the cubic case also shown (dashed line). **(D)** Exponential dependence of resistance as a function of reduced temperature. Data in A is the same as Fig. 3B for device 2 and B is the same as Fig. 1C for device 2

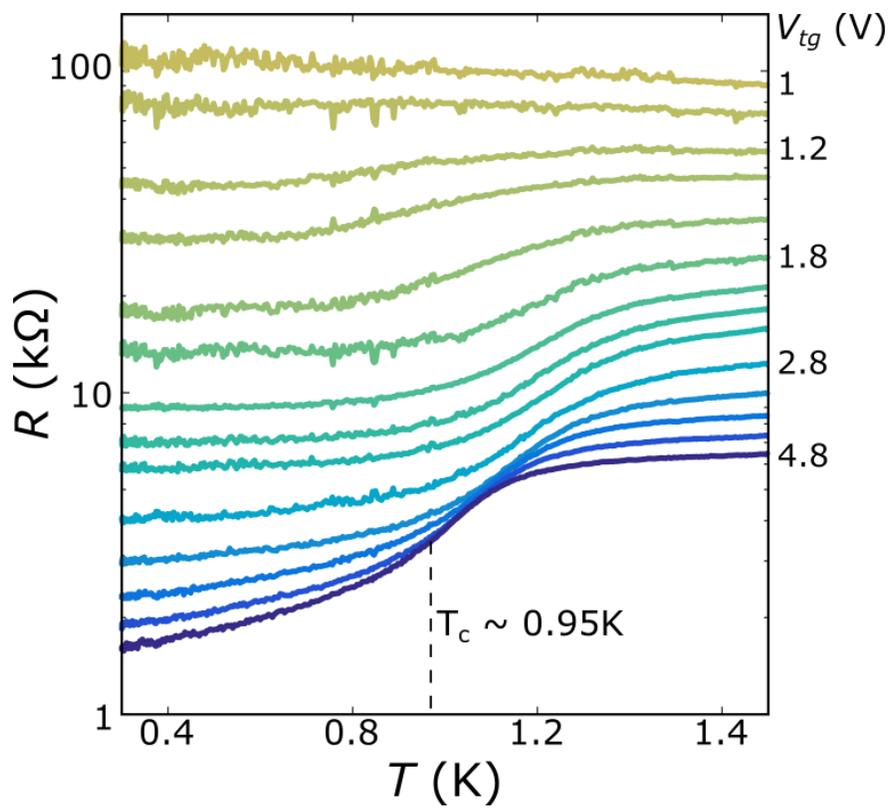

**Fig. S2 Temperature dependence for different gate voltages in device 2.** Vertical dashed line indicates $T_c$ for the largest applied gate voltage.

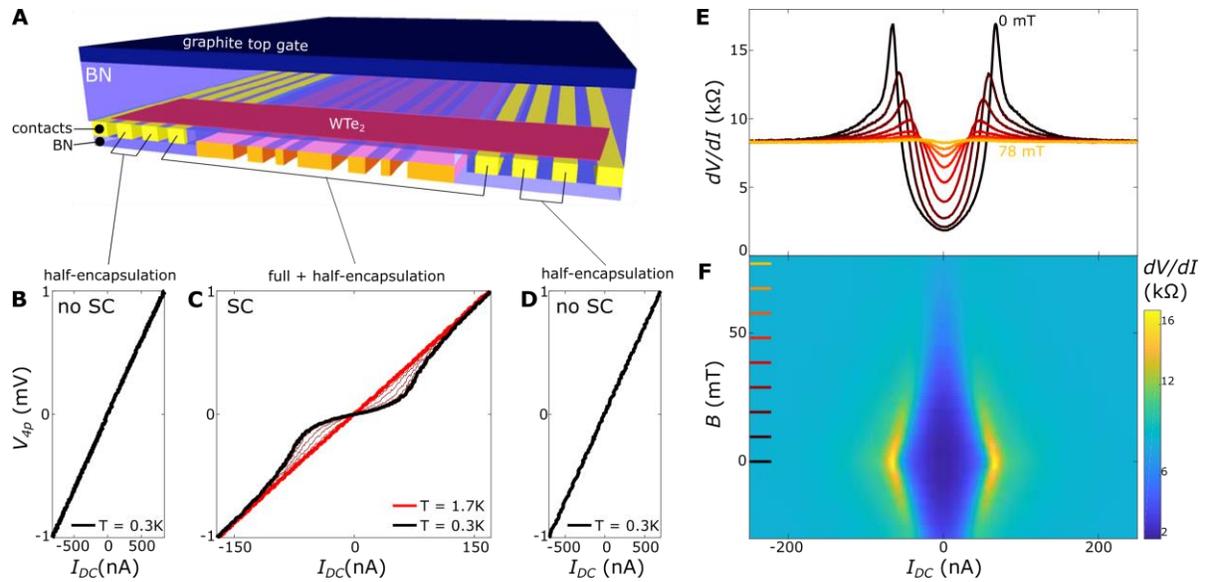

**Fig. S3 Characterization of device 2.** (**A**) Schematic of the structure of device 2. See also Ref. (*2*) for more details. B-D, Four probe V-I curves for different sets of contacts. B & D, The case of half-encapsulated $WTe_2$, in which superconductivity is not detected down to 300 mK. (**C**) The channel that includes fully encapsulated $WTe_2$ shows a transition toward superconducting behavior at low temperature. (**E**) $dV/dI$ as a function of direct current bias at different perpendicular magnetic fields (same case as panel (**C**)). (**F**) Same data as in (**E**) in color-plot form.

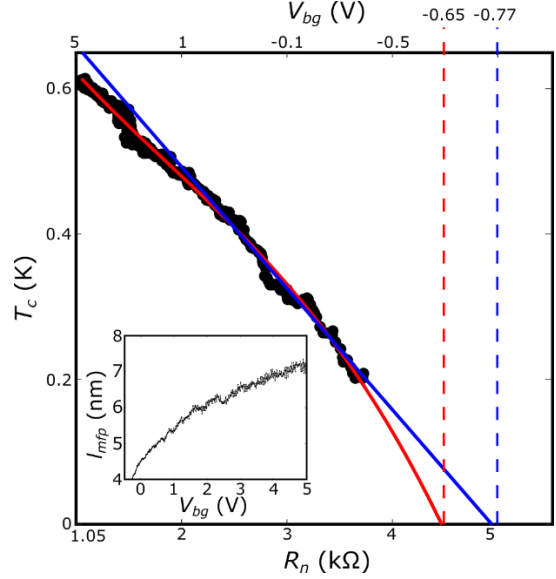

**Fig. S4 Relationship between $T_c$ and $R_n$, and gate dependence of mean free path, for device 1.** Black dots represent the experimentally measured $T_c(R_n)$ relationship. Upper x-axis shows the gate voltage for the resistances measured (note the non-linear relationship). The solid lines are phenomenological fitting curves. The blue curve is a linear fit for $R_n > 2.5$ k$\Omega$. The red curve is a cubic fit to the entire dataset. Both curves are extended to $T_c = 0$ based on the measured relationship between $V_{bg}$ and $R_n$. The critical gate voltage determined by both extrapolations are indicated by the vertical dashed lines and the corresponding labels. The red curve is used in Fig. 2B of the main text. *Inset:* Estimate for the mean free path ($l_{mfp}$) in the normal state as a function $V_{bg}$ for $V_{tg} = 5$V.

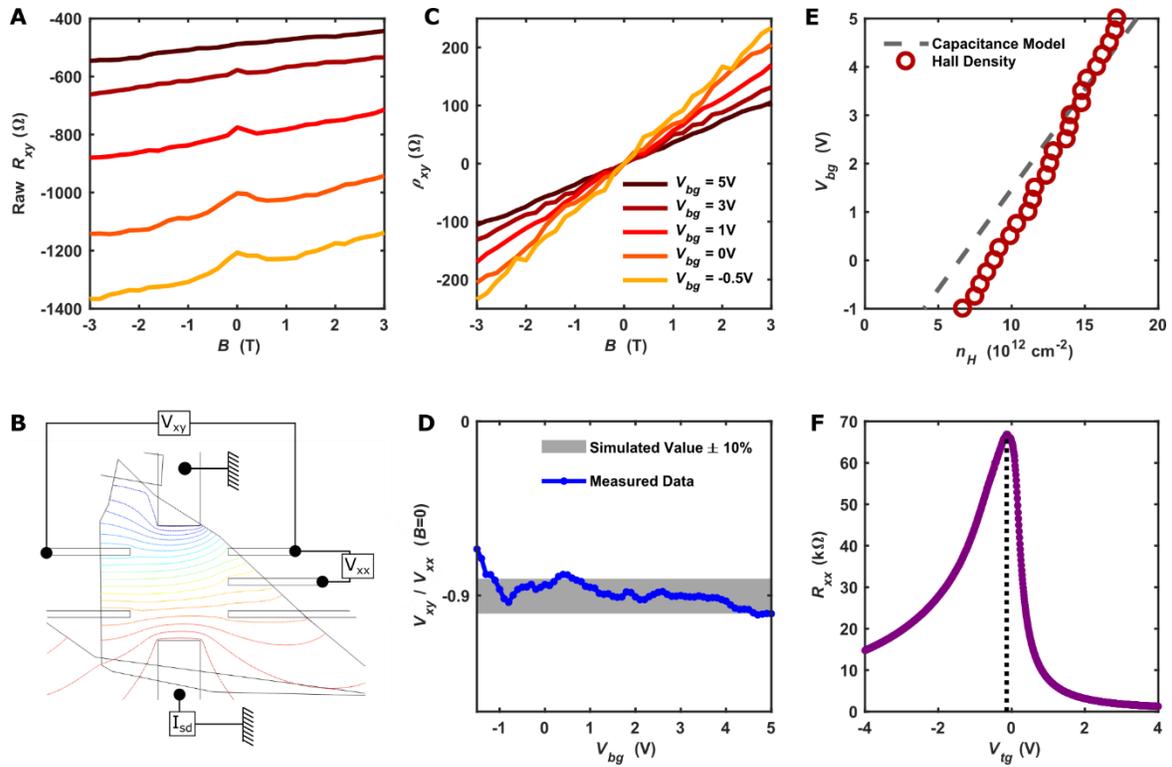

**Fig. S5 Carrier Density Analysis.** (**A**) The raw $R_{xy}$ data as a function of $B$ for five representative bottom gate voltages, taken at 2 K. $V_{tg}$ is set at 5 V. (**B**) Geometry of Device 1 for Hall effect measurements, with simulated equipotential lines drawn for current sourcing as shown. Red (purple) indicates high (low) potential. (**C**) Hall resistivity $\rho_{xy}$ obtained by antisymmetrization and correction by the simulated geometry factor for the same data as in panel (A). (**D**) Experimentally measured $V_{xy}/V_{xx}$ at zero magnetic field, in agreement with the expected value of -0.9 obtained from the simulation. (**E**) Extracted Hall density as a function of $V_{bg}$. Grey dashed line is the estimation from the capacitance model. (**F**) Longitudinal resistance as a function of top gate voltage at 90 Kelvin (with $\underline{V_{bg}}$ = 0 V). The strongly ambipolar behavior is centered near zero gate voltage.

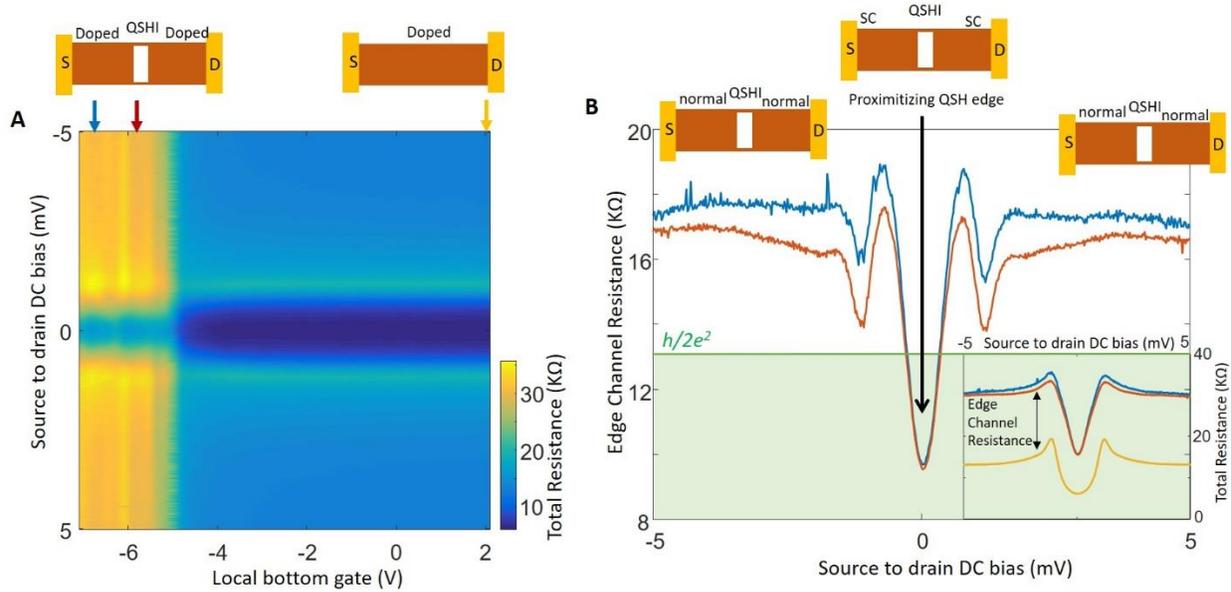

**Fig. S6 Preliminary experimental evidence suggesting the realization of superconducting helical edge modes in Device 2. (A)** Total differential resistance of the monolayer flake as a function of source to drain bias and a selected local bottom gate (with 100 nm width, device details described in Fig. S3). The monolayer is highly doped by a global top gate (3.5V) so that it will become superconducting at low temperature. The local bottom gate is used to deplete the carrier density in a short region (~100 nm in length) so that the QSH insulating state exists in that region. The two situations (entirely doped flake *vs* doped-QSH insulator-doped junctions in monolayer) are illustrated by the cartoon pictures. The data is taken at 300 mK. **(B)** *Inset:* the line cuts indicated by the arrows in A, where the local bottom gate is set to 2V (yellow), -5.8V (orange) and -6.8V (blue). The differences between the orange/blue curves and the yellow curve is plotted as the main panel in B, which measures the edge channel resistance. The cartoon pictures indicate the realized configurations in the monolayer flake.

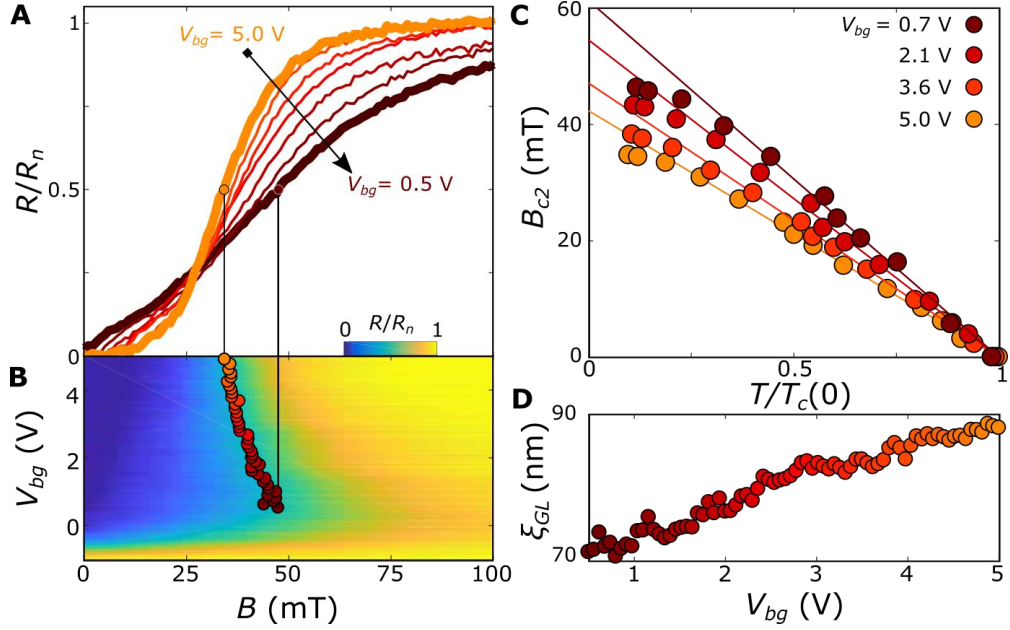

**Fig. S7 Perpendicular magnetic field effect and coherence length.** (**A**) Gate-dependent $R$ traces under the application of out-of-plane magnetic fields, normalized to the normal state resistance $R_n$. (**B**) Color map of the same data, as a function of $V_{bg}$ and $B$. Circles denote $B_{c2,\perp}$, referring to the magnetic field at which 50% of the normal state resistance is recovered. (**C**) Temperature dependence of $B_{c2,\perp}$ for selected gate voltages, including fits to the GL formula for $T$ close to $T_c$ (see main text). (**D**) Coherence length extracted from fits. Color of all data points and lines represents the gate voltage of that data point.